\documentclass[a4paper]{panl}

\usepackage{cite}
\usepackage{wrapfig}
\usepackage{graphicx}
\usepackage{amssymb}
\usepackage{amsfonts}
\usepackage{amsmath}
\usepackage{longtable}
\usepackage{rotating}
\usepackage{lscape}
\usepackage{epsfig}
\usepackage{multirow}

\originalTeX

\begin{document}

\issuearea{Physics of Elementary Particles and Atomic Nuclei. Theory}

\title{Lifetime estimations and  a non-monotonic initial energy density in heavy ion collisions at RHIC and LHC}

\maketitle

\authors{G.~Kasza $^{a,b}$ \footnote{E-mail: kasza.gabor@wigner.mta.hu}, 
T.~Cs\"org\H{o} $^{a,b,c}$ \footnote{E-mail: tcsorgo@cern.ch}}

\from{$^{a}$\,MTA Wigner FK, H-1525 Budapest 114, P.O.Box 49, Hungary}
\vspace{-3mm}
\from{$^{b}$\,EKE KRC, H-3200 Gy\"ongy\"os, M\'atrai \'ut 36, Hungary}
\vspace{-3mm}
\from{$^{c}$\,CERN, CH-1211 Geneva 23, Switzerland}

\begin{abstract}
We highlight some  connections between the final state hadronic observables and the initial conditions using a recently found new exact family of solutions of relativistic hydrodynamics. These relations provide explicit examples of the scaling behaviour in relativistic hydrodynamics and may provide an advanced estimate of the lifetime and the initial energy density in
$\sqrt{s_{NN}} = 62.4$, $130$, and $200$ GeV Au+Au collisions at RHIC and $\sqrt{s_{NN}} = 5.0$ TeV Pb+Pb and $5.44$ TeV Xe+Xe as well as $\sqrt{s} = 7$, $8$ and $13$ TeV p+p collisions at LHC energies. A surprising result is that these advanced
estimates yield a non-monotonic increase of the initial energy density with increasing collision energy at the RHIC energy range.
\end{abstract}

\vspace*{6pt}

\noindent

%PACS:

\vspace*{12pt}

\label{sec:intro}
\section*{\it Introduction}
Recently, we have published a series of manuscripts that presents a new family of exact solutions of relativistic hydrodynamics~\cite{Csorgo:2018pxh},
and its applications to the evaluation of the pseudorapidity distributions~\cite{Csorgo:2018fbz}, the longitudinal HBT radii~\cite{Kasza:2018jtu},
the estimation of the initial energy densities~\cite{Csorgo:2018crb}
as well as the application of these results to the analysis of experimental data at RHIC and LHC energies~\cite{Kasza:2018qah}.
In this conference contributions we highlight some of the most beautiful results, that are given in full detailed in refs.~\cite{Csorgo:2018pxh,Csorgo:2018fbz,Kasza:2018jtu,Csorgo:2018crb,Kasza:2018qah}.

\section*{\it New, exact solutions of relativistic, perfect fluid hydrodynamics}
The equations of relativistic perfect fluid hydrodynamics are given in terms of
the entropy density, denoted by $\sigma$, the four velocity  $u^{\mu}$, normalized as $u^{\mu}u_{\mu}=1$, and the energy-momentum four tensor $T^{\mu \nu}$, as follows. These fields depend on  $x\equiv x^{\mu} = (t,r_x,r_y,r_z)$. 
%governed by the continuity equation of entropy density and the conservation of the four-momentum:
\begin{eqnarray}
	\partial_{\mu}\left(\sigma u^{\mu}\right)&=&0, 
	\label{e:entropy} \\
	\partial_{\nu}T^{\mu \nu} &= &0.
	\label{e:energy-momentum} 
\end{eqnarray}

For perfect fluids, the energy-momentum four tensor is 
\begin{equation} 
	T^{\mu \nu}=\left(\varepsilon+p\right)u^{\mu}u^{\nu} - pg^{\mu \nu},
	\end{equation}
where $\varepsilon \equiv \varepsilon(x)$ is the energy density and $p \equiv p(x)$ is the pressure, and $g^{\mu \nu} =\mbox{\rm diag}(1,-1,-1,-1)$ stands for the Minkowskian metric. The above set of equations
%set of equation provides 5 equations for 6 unknown fields. 
is closed by the following equation of state:
%
%closes and makes possible to solve this system of equations. 
%In this manuscript, the following equation of state is assumed:
	\begin{equation}
	p = \bar{c}_s^2 \varepsilon = \varepsilon/\kappa,
	\end{equation}
where $\bar{c}_s\equiv 1/\sqrt{\kappa}$ is a temperature independent, average value 
of the speed of sound, that was measured by the PHENIX collaboration in ref.~\cite{Adare:2006ti}: $\bar{c}_s=0.35\pm 0.05$, corresponding to $\kappa=10^{+1}_{-3}$.
	\begin{figure}
	\centering
	\includegraphics[scale=0.45]{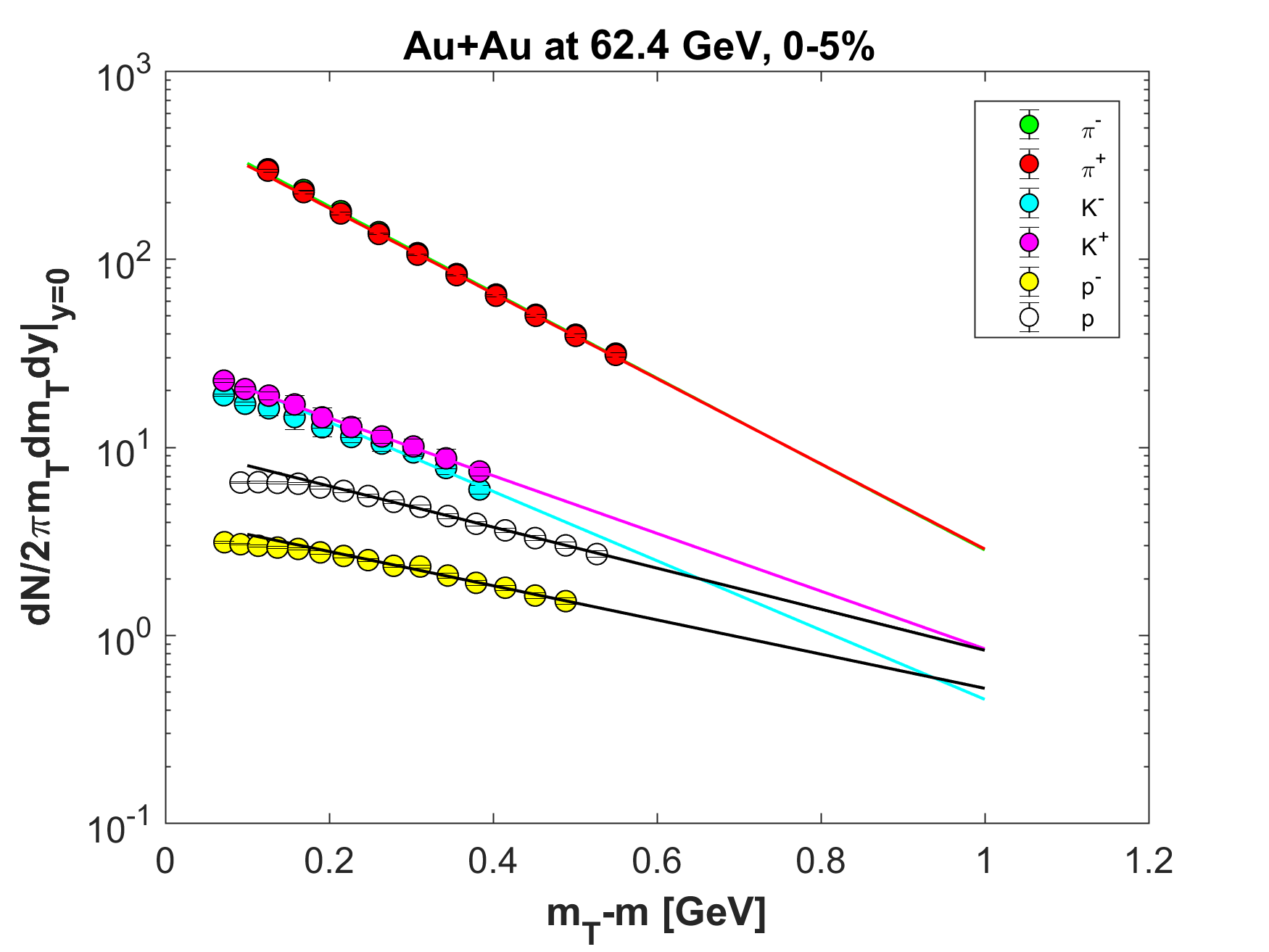}
	\includegraphics[scale=0.45]{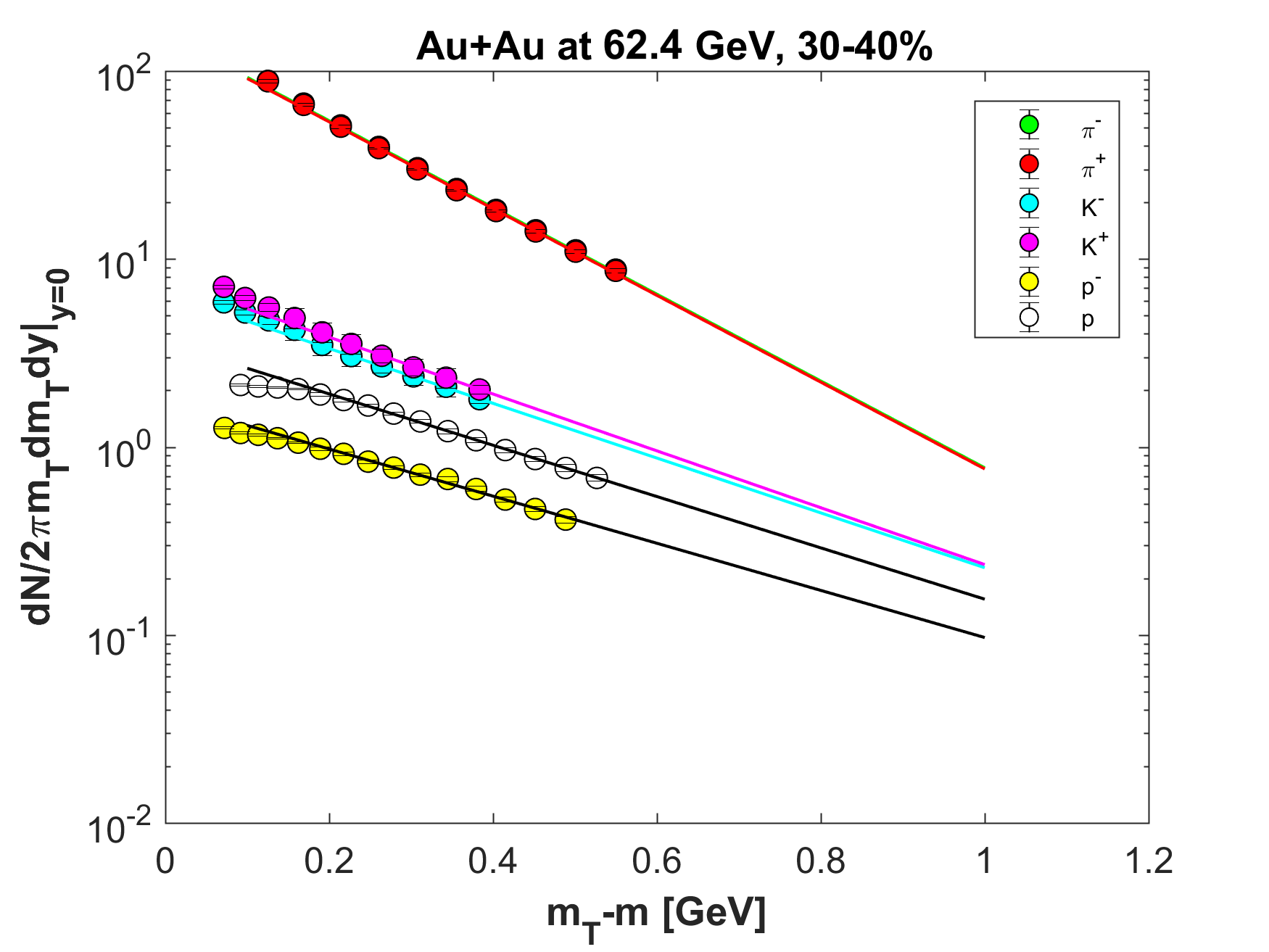}
	\includegraphics[scale=0.45]{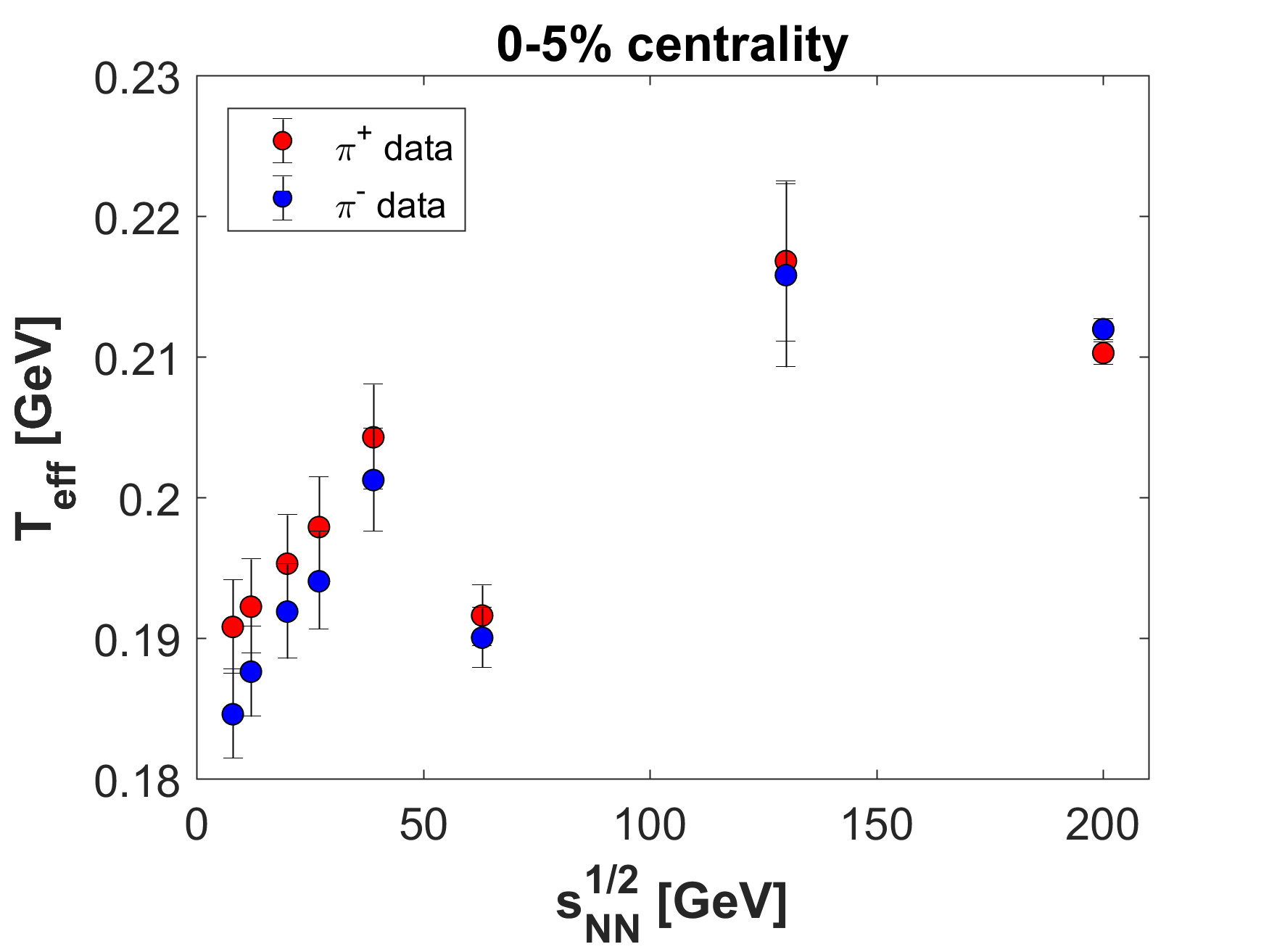}
	\includegraphics[scale=0.45]{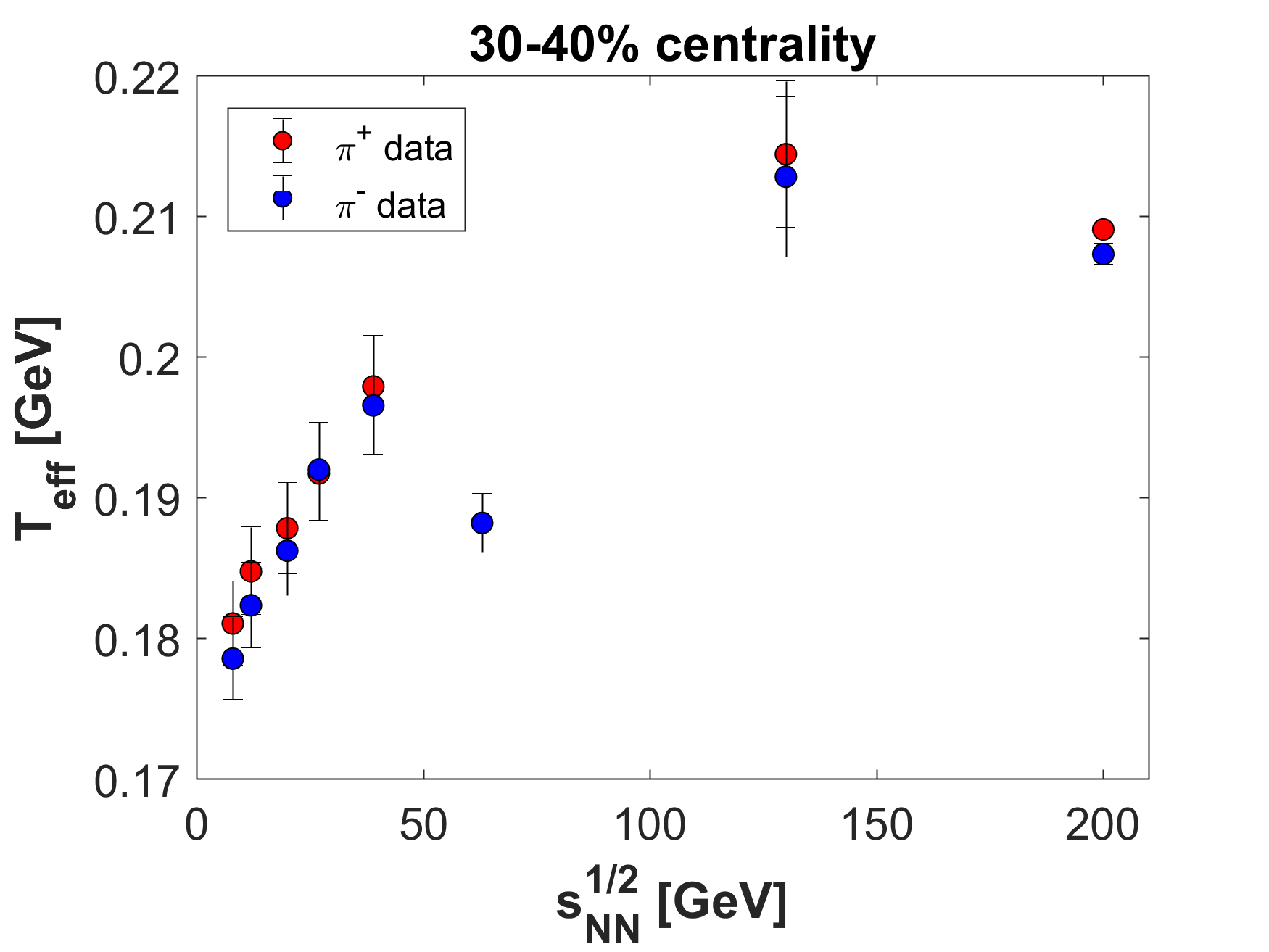}
 	\caption{The top left panel indicates the result of fits with $A \exp\left[-(m_T -m) / T_{\rm eff}\right]$ to the invariant  momentum distribution of
	$\pi^{\pm}$, $K^{\pm}$, $p$ and $\overline{p}$ to $\sqrt{s_{NN}} = 62.4 $ GeV Au+Au collisions in the $0 - 5 $ \% centrality class.
	The top right panel shows the same in the $30 - 40$ \% centrality class. Bottom left panel shows the excitation function of the effective temperature of positively and negatively charged pions in the $0 - 5 $ \% centrality class. Bottom right panel shows  the same in the $30 - 40$ \% centrality class. The bottom panels are indicating a non-monotonic behaviour of the pion slope parameters with increasing energy, 
	%with a sudden decrease of the slopes at around $\sqrt{s_{NN}}$  $=$ $62.4$ GeV 
	in both centrality classes. }
	\label{fig:IS+pt_spectra}
	\end{figure}
An exact and analytic, finite and accelerating, 1+1 dimensional solution of relativistic perfect fluid hydrodynamics was recently found by Cs\"org\H{o}, Kasza, Csan\'ad and Jiang (CKCJ)~\cite{Csorgo:2018pxh} as a family of parametric curves, obtained by using the Rindler-coordinates:
$	\left(\tau,\eta_x\right) =  
	\left(\, \sqrt{t^2-r_z^2}\, ,           
	\frac{1}{2}\textnormal{ln}\left[\frac{t+r_z}{t-r_z}\right]\,\right),
$
where $\tau$ stands for the longitudinal proper time and $\eta_x$ is the space-time rapidity. 
The four-velocity is chosen as $u^{\mu}= \left(\cosh\left(\Omega\right),\sinh\left(\Omega\right)\right)$ and it was  assumed that the fluid rapidity $\Omega \equiv \Omega (\eta_x)$ is independent of the proper time. The new class of CKCJ solutions was presented in  ref.~\cite{Csorgo:2018pxh} and the corresponding pseudorapidity distributions were obtained as parametric curves in ref.~\cite{Csorgo:2018fbz}. 

We present this formulae, as well as an advanced  
estimation of the initial energy densities~\cite{Csorgo:2018crb},
and the results on the longitudinal HBT radii of ref.~\cite{Kasza:2018jtu},
as derived from the CKCJ exact solution of relativistic hydrodynamics.
%	\begin{eqnarray}
%	\eta_x(H)  & = & \Omega(H) -H, 
%	\label{e:etaH}\\ 
%	\Omega(H)  & = & 
%	\frac{\lambda}{\sqrt{\lambda-1}\sqrt{\kappa-\lambda}}
%	\textnormal{arctan}\left(\sqrt{\frac{\kappa-\lambda}
%		{\lambda-1}}\textnormal{tanh}\left(H\right)\right), 
%	\label{e:OmegaH} \\ 
%	\sigma(\tau,H)&= & \sigma_0 
%	\left(\frac{\tau_0}{\tau}\right)^{\lambda}
%	\mathcal{V}_{\sigma}(s) \left[1+\frac{\kappa-1}{\lambda-1}
%	\textnormal{sinh}^2(H)\right]^{-\frac{\lambda}{2}},
%	\label{e:sigmasol} \\
%	T(\tau,H)  & = & T_0 
%	\left(\frac{\tau_0}{\tau}\right)^{\frac{\lambda}{\kappa}} 
%	\mathcal{T}(s) 
%	\left[1+\frac{\kappa-1}{\lambda-1}\textnormal{sinh}^2(H)
%	\right]^{-\frac{\lambda}{2\kappa}},
%	\label{e:Tsol}\\ 
%	\mathcal{T}(s) & = & 
%	\frac{1}{\mathcal{V}_{\sigma}(s)},
%	\label{e:scalingsol} % \\
%	\qquad\quad
%	s(\tau,H) \, = \,
%	\left(\frac{\tau_0}{\tau}\right)^{\lambda-1} 
%	\textnormal{sinh}(H)\left[1 + \frac{\kappa-1}{\lambda-1}
%	\textnormal{sinh}^2(H)\right]^{-\lambda/2}.
%	\label{e:sH}
%	\end{eqnarray}	
This solution is an explicit function of the longitudinal proper-time $\tau$, while its the dependence on the space-time rapidity $\eta_x$ is
given in terms of parametric curves, parameterized in terms of $H = \Omega - \eta_x$.
These solutions are constrained to a cone inside the forward lightcone, limited by constant pseudorapidity lines, see ref.~\cite{Csorgo:2018pxh} for details.

%The normalization constants $\sigma_0$ and $T_0$ stand for $\sigma(\tau_0,H=0)$ and $T(\tau_0,H=0)$, where $\tau_0$ is the initial proper time. 

\begin{figure}
	\centering
	\includegraphics[scale=0.45]{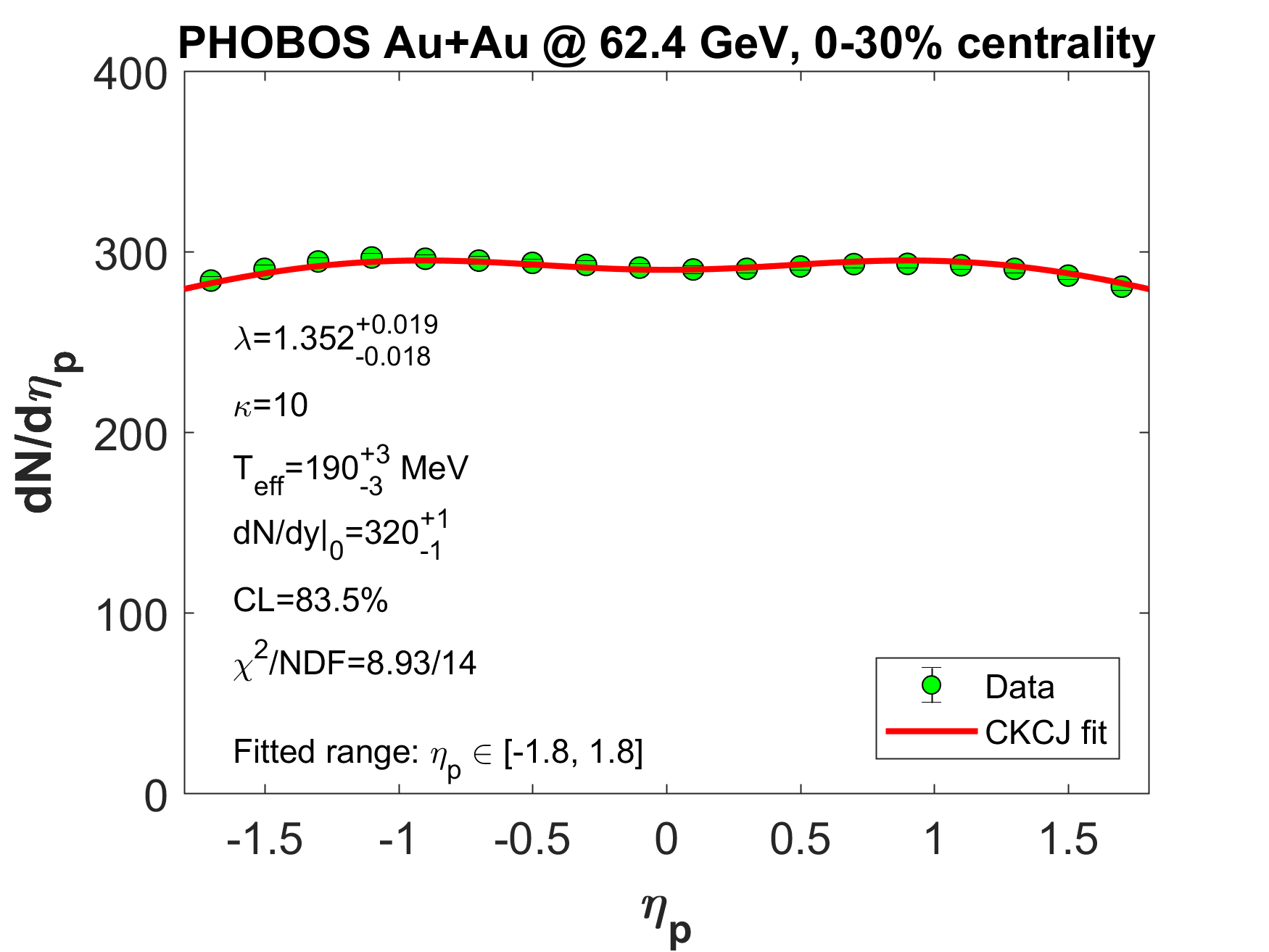}
	\includegraphics[scale=0.45]{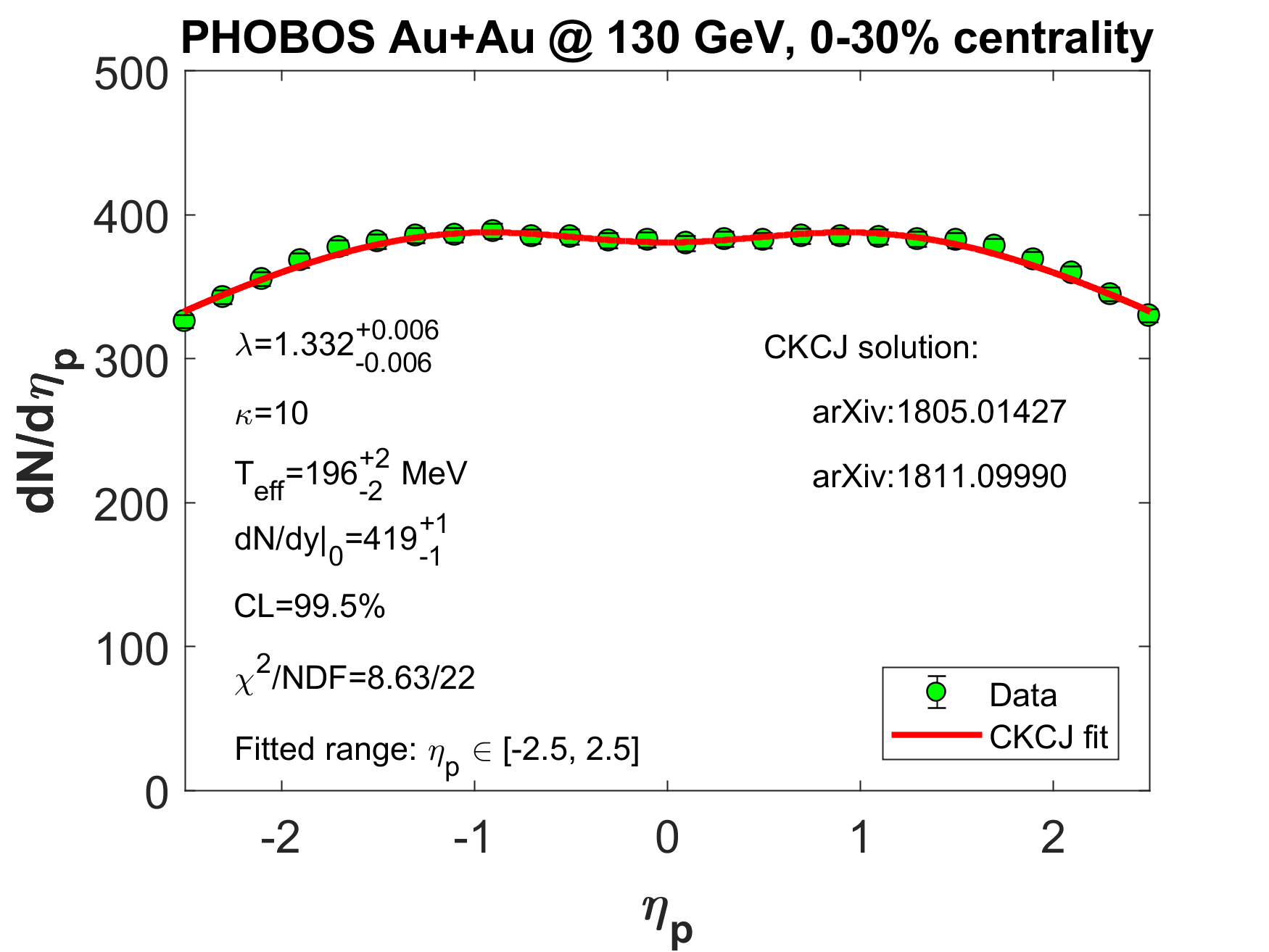}
	\caption{Description of PHOBOS results~\cite{Alver:2010ck} on the pseudorapidity distributions of $\sqrt{s_{NN}}$ $=$ $62.4$ and $130$  GeV, 0-30 \%  Au+Au collisions with the CKCJ solutions of relativistic hydrodynamics, using eqs.~(\ref{e:dndeta-function}-\ref{e:y-eta}).
	Details of the fitting method are described in ref.~\cite{Kasza:2018qah}.}
	\label{fig:dndeta}
	\end{figure}

	\begin{figure}
	\centering
	\includegraphics[scale=0.6]{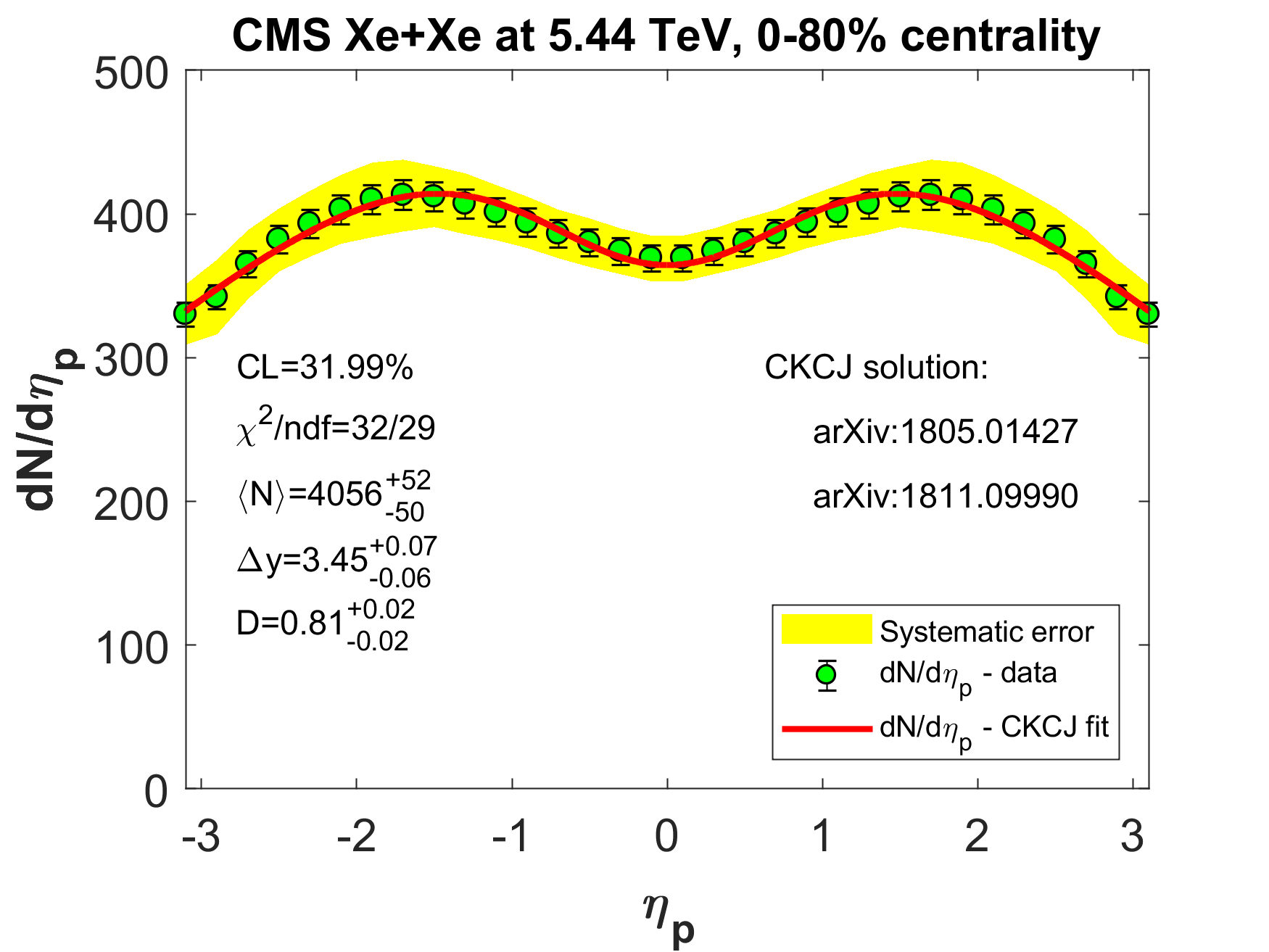}
	\caption{Fits of the pseudorapidity density with the CKCJ hydro solution~\cite{Csorgo:2018pxh}, to CMS Xe+Xe data at $\sqrt{s_{NN}} = $ 5.44 TeV \cite{Sirunyan:2019cgy} in the 0-80 \% centrality class, using eqs.~(\ref{e:dndeta-function},\ref{e:dndy-Gauss-width},\ref{e:y-eta})~\cite{Kasza:2018qah}. The fit parameters are the mean multiplicity $\langle N\rangle$, the Gaussian width of 
	%pseudorapidity distribution 
	$\Delta y$ and the dimensionless dip parameter $D=m/\overline{p}_T$. Yellow band stands for the systematic errors of ref.~\cite{Sirunyan:2019cgy}.}
	\label{fig:xexe_dndeta}
	\end{figure}

\section*{\it Pseudorapidity density and hydrodynamic scaling behaviour}
Recently, a robust functional form has been derived~\cite{Kasza:2018qah} from the CKCJ solution~\cite{Csorgo:2018pxh}:
%of relativistic hydrodynamics:
\begin{eqnarray}
	\frac{dN}{d\eta_p} & \approx & 
	\frac{\langle N\rangle}{\big(\displaystyle\strut 2\pi\Delta^2 y\big)^{1/2}}%{\frac{1}{2}}}
	\frac{\cosh(\eta_p)}{\big[\displaystyle\strut  (m/\bar{p}_T)^2 
	                        + \cosh^2(\eta_p) \big]^{1/2}}%{\frac{1}{2}} } 
	\exp\Bigg(-\frac{y^2}{ 2 \Delta^2 y \phantom{\Big|} }  \Bigg) \,
	\Bigg|_{y = y(\eta_p)}, \label{e:dndeta-function} \\
		\frac{1}{\Delta^2 y} & = & (\lambda - 1)^2 \, \left[ 1 + \left( 1 - \frac{1}{\kappa} \right) 
	\left(\frac{1}{2} + \frac{m}{T_{\rm eff}} \right) \right] , 
	\label{e:dndy-Gauss-width}
	\\
	y(\eta_p) & \approx & \tanh^{-1} \left( \sinh(\eta_p) / \sqrt{ (m/\bar{p}_T)^2 + \cosh^2(\eta_p) } \right) .\label{e:y-eta}
	\end{eqnarray}	
This formula can be used to fit the
pseudorapidity density distributions at RHIC and LHC energies from proton-proton to (symmetric) heavy ion collisions near mid-rapidity,
outside the spectator fragmentation regions, where possible shock-wave effects might have to be taken into account too.
Here parameter $\lambda$ is a constant of integration, that controls the width of the pseudorapidity distribution.
It can be determined from fits to the pseudorapidity distribution data.
Given that $\Delta y $ is a physical fit parameter, that corresponds to a combination of the equation of state parameter $\kappa$, the mass $m$  of the
particles (predominantly pions),  and $T_{\rm eff}$ is the slope parameter of the single-particle spectra, eq.~(\ref{e:dndy-Gauss-width}) is a beautiful
example of the hydrodynamical scaling behaviour: different hydrodynamical sources may lead to the same pseudo-rapidity distributions,
if the relevant combination $\Delta y$ of the hydrodynamical parameters $\lambda$, $\kappa$, $m$ and $T_{\rm eff}$ is the same. 
The boost-invariant Hwa-Bjorken solution corresponds to the $\lambda \rightarrow 1$ limit.
The depression of the pseudorapidity distribution at mid-rapidity is controlled by the 
average transverse momentum at mid-rapidity, denoted by $\bar{p}_T$, which  together with $T_{\rm eff}$ can also be determined from fits to the transverse mass 
spectra as indicated on Fig.~\ref{fig:IS+pt_spectra}. If $\lambda -1 \ll 1$, $\bar{p}_T = \sqrt{(m + T_{\rm eff})^2 - m^2}$, see ref.~\cite{Kasza:2018qah} for more
details. The above form of the pseudorapidity distribution is  normalized to $\langle N \rangle$,
the mean multiplicity.  In this case the normalization 
should also be changed from the infinite mean multiplicity and the diverging $\Delta y$ to their ratio, the finite mid-rapidity density,
given by $\langle N\rangle /( 2\pi\Delta^2 y )^{1/2}$.
%constants $\sigma_0$ and $T_0$ stand for $\sigma(\tau_0,H=0)$ and $T(\tau_0,H=0)$, where $\tau_0$ is the initial proper time. 
	\begin{figure}
	\centering
	\includegraphics[scale=0.45]{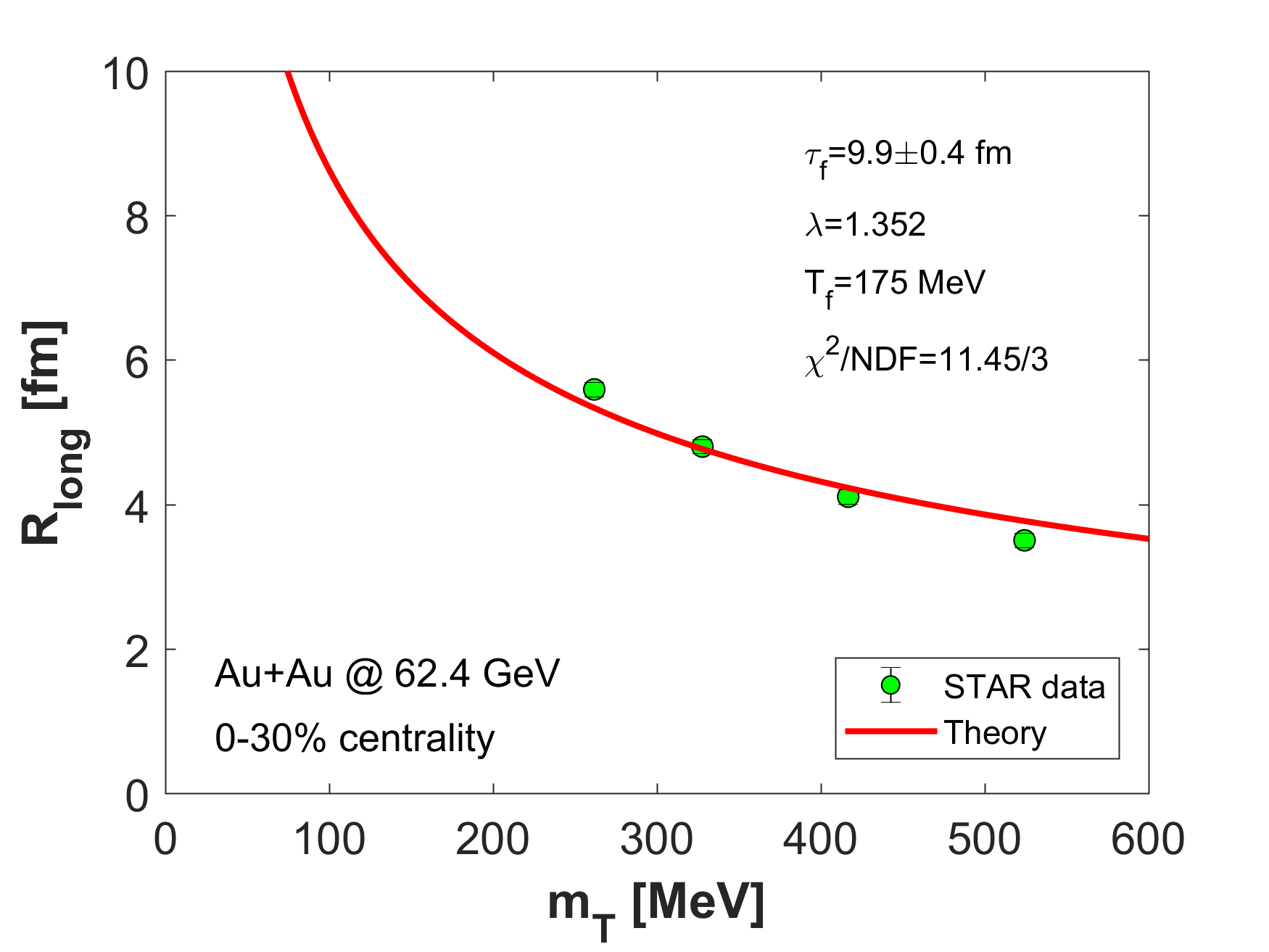}
	\includegraphics[scale=0.45]{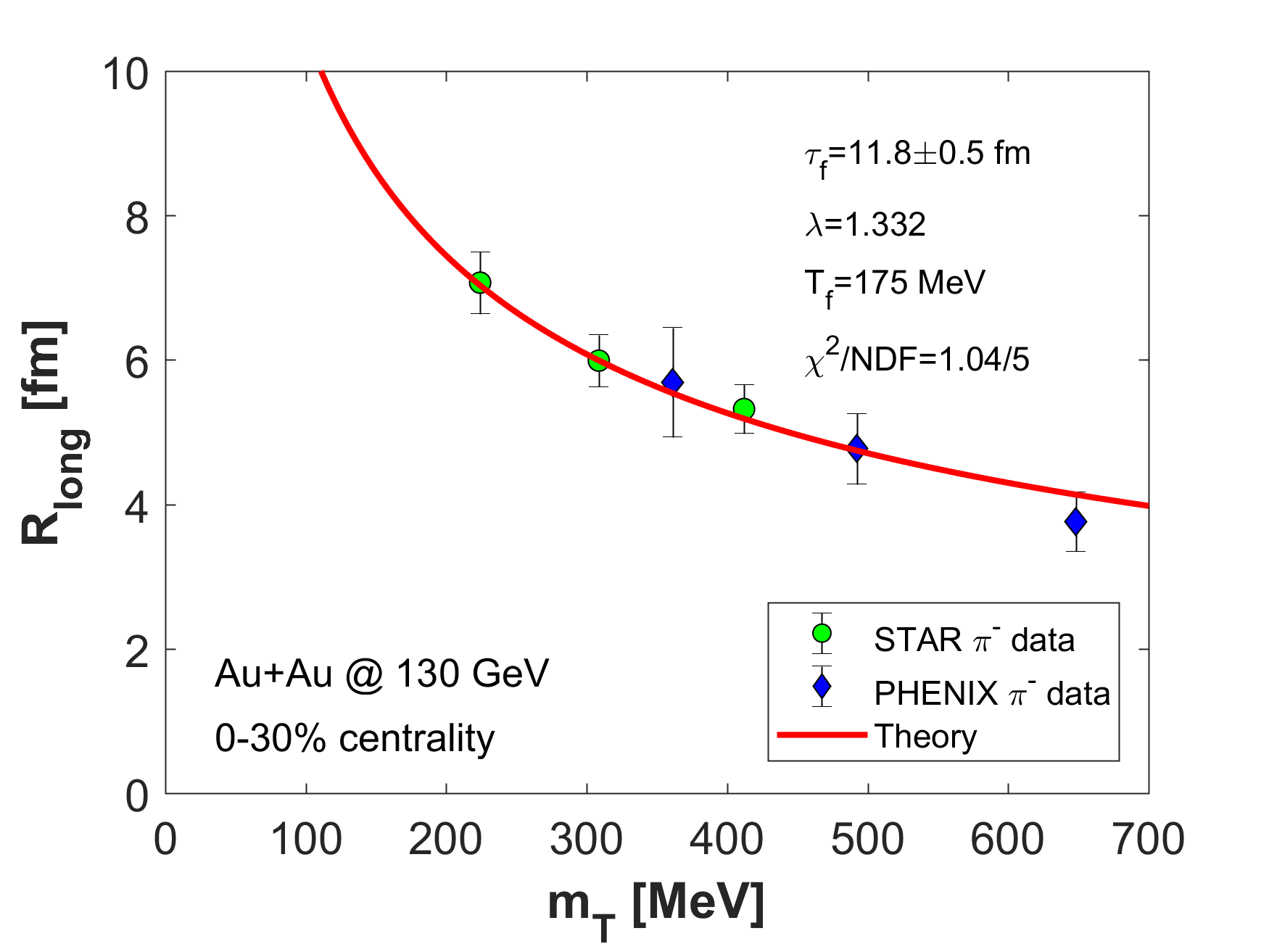}
	\caption{Fits of the longitudinal HBT-radii with the CKCJ hydro solution~\cite{Csorgo:2018pxh}, to STAR Au+Au data at $\sqrt{s_{NN}} = $ 62.4 GeV~\cite{Adamczyk:2014mxp} (left) and to PHENIX and STAR Au+Au data at $\sqrt{s_{NN}} = $ 130 GeV~\cite{Adcox:2002uc} (right) in the 0-30 \% centrality class, for a fixed centrality class and colliding system.}
	\label{fig:rlong}
	\end{figure}	

The CKCJ solution was shown to describe the pseudorapidity density distribution of $p+p$ collisions at $\sqrt{s} = 7$ and $8$ TeV
in ref.~\cite{Csorgo:2018pxh}. The description of the pseudorapidity distribution of 40 - 50 \%  Pb+Pb collisions at $\sqrt{s_{NN}} = 5$ TeV was given in 
ref.~\cite{Csorgo:2018fbz}, while fits to $Au$ + $Au$ collision  data at $\sqrt{s_{NN}} = 200$ GeV were shown in ref.~\cite{Kasza:2018jtu}. Several other plots were shown in conference presentations that indicate that the CKCJ solution of relativistic hydrodynamics
describes well the pseudorapidity distributions from $p+p$ throuhg $Xe+Xe$ and $Au+Au $ to  $Pb+Pb$ collisions, from the colliding energies
of $\sqrt{s_{NN}} = 20$ GeV to the presently largest LHC energy of $\sqrt{s} = 13$ TeV. Most recently, we have described the CMS measurement of the
pseudorapidity density in $Xe+Xe$ collisions at $\sqrt{s_{NN}} = $ 5.44 TeV \cite{Sirunyan:2019cgy} in the 0-80 \% centrality class,
as indicated on Fig.~\ref{fig:xexe_dndeta}.

\section*{\it Life-time and initial energy density estimation at RHIC energies}
In this section we present a new method, which is developed to determine the initial energy density of the expanding fireballs, as a function of only one parameter, the initial proper time. These method can be summarized as follows:
{\it (i) } Evaluate the effective temperature ($T_{\rm eff}$) from fits to the invariant momentum distribution of identified hadrons, as a function of $m_T-m$.
{\it (ii)} Determine the acceleration parameter $\lambda$ from a fit with  eq.~(\ref{e:dndeta-function}), by using eqs.~(\ref{e:dndy-Gauss-width},\ref{e:y-eta}) from fits to measured pseudorapidity density data.  See refs.~\cite{Csorgo:2018pxh, Csorgo:2018fbz, Kasza:2018qah} for details.
{\it (iii)} Fit the $m_T$ dependence of the londitudinal HBT-radii   $R_{\rm long}$, to evaluate the life-time parameter of the medium. 
This HBT radius parameter was derived from the CKCJ solution in ref.~\cite{Csorgo:2018crb} as follows:
\begin{equation}
	R_{\rm long}=\tau_f \Delta \eta_x \approx \frac{\tau_f}{\sqrt{\lambda\left(2\lambda-1\right)}} \sqrt{\frac{T_f}{m_T}},
	\label{e:Rlong-ckcj}
\end{equation}
where $\tau_f$ is the life-time parameter, and $T_f$ stands for the kinetic freeze-out temperature.
{\it (iv) } Finally, one can use the fitted parameters to evaluate the initial energy density by our new formula that was calculated exactly from the CKCJ solution. It corrects the Bjorken estimation by taking into account the non boost-invariant expansion and the finite width of the 
pseudo-rapidity distribution, as represented by $\lambda \neq 1$, as well as  the work, done by the pressure during the fireball evolution \cite{Kasza:2018jtu}:
	\begin{equation}\label{eq:IED-CKCJresult}
	\varepsilon_0(\kappa,\lambda)=\varepsilon_0^{\rm Bj} \left(2\lambda-1\right)\left(\frac{\tau_f}{\tau_0}\right)^{\lambda\left(1+\frac{1}{\kappa}\right)-1},
	\end{equation}	
	where $\varepsilon_0^{\rm Bj}$ is Bjorken's estimate and it can be expressed as \cite{Bjorken:1982qr}:
	\begin{equation}\label{eq:IED-bjorken}
	\varepsilon_0^{\rm Bj}=\frac{\langle E_T \rangle}{S_\perp \tau_0}\left.\frac{dN}{d\eta_p}\right|_{\eta_p=0}.
	\end{equation}	
	In this Bjorken estimate, $\langle E_T \rangle$ is the average thermalized, transverse energy and $S_\perp$ stands for the overlap area of the colliding nuclei.
%\end{enumerate}

We have gone through on these steps, so we estimated the initial energy density of RHIC Au+Au collisions at 3 different colliding energies, namely: $\sqrt{s_{NN}}=$62.4, 130 and 200 GeV. Here, we can only highlight one of the surprizing results,
which indicate an unexpected feature of the strongly interacting quark-gluon plasma (sQGP) created in Au+Au collisions at RHIC energies. 
Our advanced initial energy density estimate of $\sqrt{s_{NN}}=$ $130 $ and $200$ GeV collisions is detailed in ref.~\cite{Kasza:2018qah}.
%including also a lower bound on the initial energy density in $\sqrt{s_{NN}}=$ $5$ TeV $Pb+Pb$ collisions as well.

\begin{figure}
	\centering
	\includegraphics[scale=0.7]{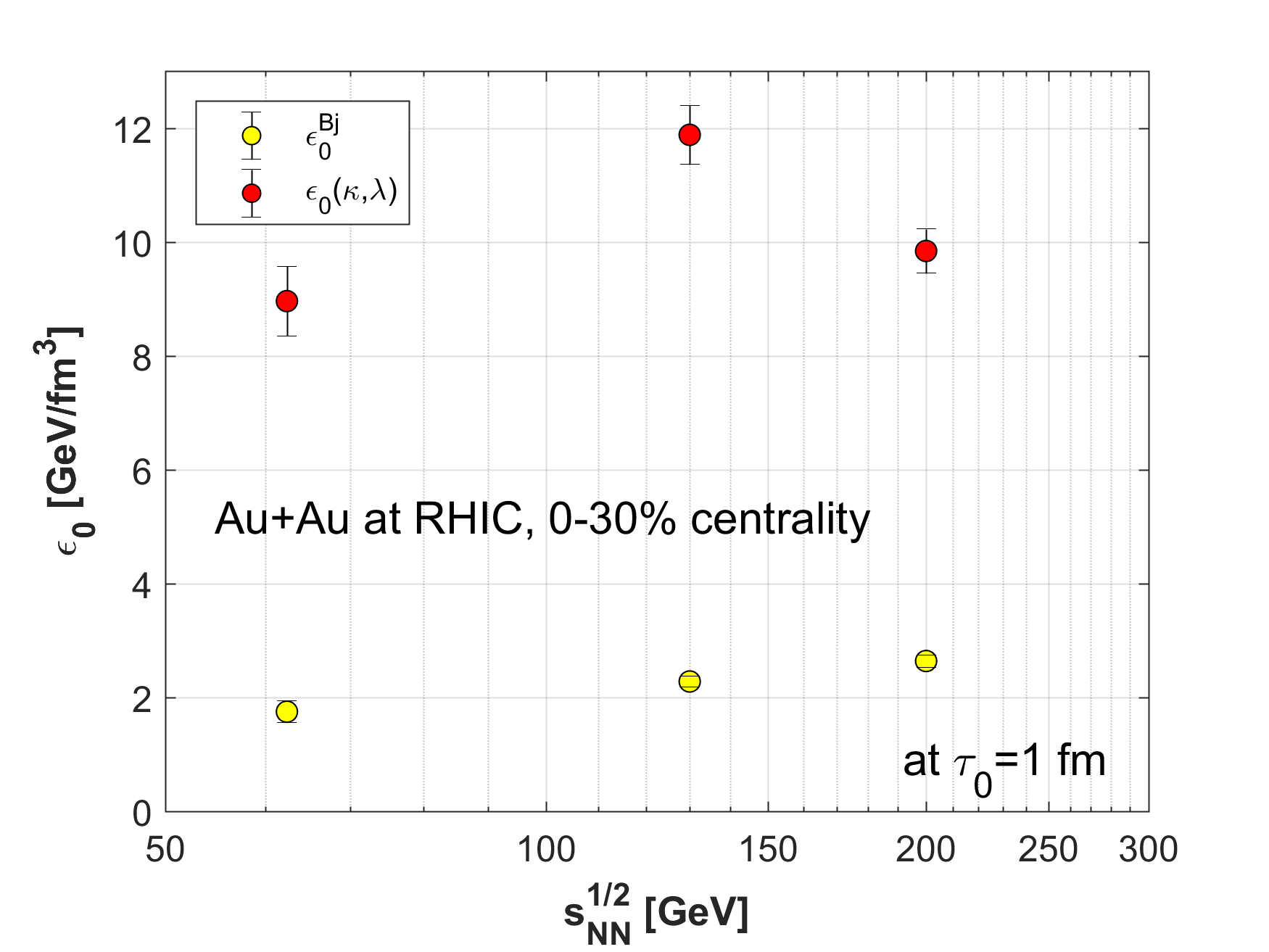}
	\caption{Initial energy density estimations as a function of the colliding energy. The calculations were performed at 62.4 GeV, 130 GeV and 200 GeV. The red dots stand for the results of our advanced initial energy estimate, based on eq.~\eqref{eq:IED-CKCJresult}, while the yellow dots correspond to Bjorken's 
	estimate. Our advanced estimates indicate a surprising and  non-monotonic behavior of the initial energy density as a function of the collision energy.
	}
	\label{fig:ied_nonmon}
	\end{figure}

Fig.~\ref{fig:ied_nonmon} indicates the initial energy densities as a function of the colliding energies in the 0-30 \% centrality class. 
This figure immediately highlights the surprising feature we have just mentioned.
Although Bjorken's estimate suggests a monotonic increase of the initial energy density with increasing energy, our advanced
initial energy density estimate indicates a non-monotonic behavior.

This result is to be considered as preliminary and treated carefully, due to the limitations of the CKCJ solution:
this solution does not include transverse dynamics, and the temperature dependence of the speed of sound is replaced by an average value.
Both shear and bulk viscosity effects are neglected. However, we have cross-checked in ref~\cite{Kasza:2018qah},
that our analytical results on the initial energy density  yields similar time evolution for the energy density in the center of the fireball
to that of an 1+3 dimensional numerical solution of the equations of relativistic hydrodynamics using the lattice QCD equation of state~\cite{Bozek:2009ty}. 
This comparison provided  a surprisingly good agreement, suggesting that the non-monotonic  behavior of the initial energy density with increasing energy 
of the RHIC beam energy scan may be a robust feature of the data. Of course, it is conceivable that the match is a coincidence, since ref.~\cite{Bozek:2009ty} 
did not made explicit the centrality class of the calculation. In addition, of course, the question may arise about the initial energy density of LHC energies predicted by the CKCJ solution. We have also estimated with our new method the lower limit of the initial energy density of $Pb$+$Pb$ collisions at 
$\sqrt{s_{NN}} = 2.76$ TeV, in the 10-20 \% centrality class. The result
%, published in ref.~\cite{Kasza:2018qah}, 
satisfied our expectation, 
namely the lower limit of the initial energy density at such a high colliding energy is higher than our initial energy density estimate  at $\sqrt{s_{NN}} = 200$ GeV,
the top RHIC energy for $Au$+$Au$ collisions.

{\it In summary}, we highlighted in this work two of our remarkable theoretical results from ref.~\cite{Kasza:2018qah} :

{\it (i)} A simple and beautiful formula was found to describe the pseudorapidity
density distribution from the CKCJ solution of relativistic hydrodynamics.
This formula is given by eqs. (\ref{e:dndeta-function}-\ref{e:y-eta}), 
and it describes not only $p$ + $p$ and heavy ion data at RHIC and LHC energies, 
but it also describes excellently the recent CMS data on Xe + Xe collisions exceedingly well.

{\it (ii)} We outlined our new method to extract the initial energy density of high
energy proton-proton and heavy ion collisions, that corrects  Bjorken’s oversimplified initial
ener\-gy density estimate for realistic pseudorapidity density distributions and for taking into account
the work done by the non-vanishing pressure during the expansion.

The excitation function of the initial energy density is summarized in Fig.~(\ref{fig:ied_nonmon}),
indicating a non-monotonic behaviour.
These advanced  estimates of the initial energy density may thus 
become a new tool to search for the critical point of the QCD phase diagram: in the vicinity
of the QCD critical point, several quantities may behave in a non-monotonic manner,
including life-time related observables, such as the estimated initial energy density.
Indeed, a non-monotonic behaviour of the HBT-radii has been observed in the RHIC beam energy scan, 
pointing to a QCD critical point near $\mu_B \approx 95$ MeV, corresponding to 
$\sqrt{s_{NN}} $ $\approx$  $47.5$ GeV~\cite{Lacey:2014wqa}.
Our data analysis related to the estimations of the initial energy density of Au+Au collisions
at RHIC supports independently the possibility of this kind of a scenario. This scenario is also supported by the rather robust
nature of Fig.~\ref{fig:IS+pt_spectra}, that indicates a non-monotonic dependence of the slope parameters of the single-particle
transverse mass spectra on $\sqrt{s_{NN}}$ in a similar energy range.

Further, more detailed studies are necessary to 
investigate possible shock-wave effects at lower colliding energies, together with the effects 
arising from a possible proper-time dependence of the acceleration parameter
$\lambda$ and from using a more realistic, 1+3 dimensional expansion, the  temperature
dependence of the speed of sound and using a lattice QCD equation of state.

%%%%%%%%%%%%%%%%%%%%%%%%%%%%%%%%%%%%%%%%%%

\newpage
\section*{\it Acknowledgments}
This research has been partially supported by the NKIFH grants No. FK-123842 and FK-123959,  the  EFOP 3.6.1-16-2016-00001 grant (Hungary),
and  THOR, the EU COST Action CA15213.

\end{document}